\documentclass[a4paper,11pt]{article}
\pdfoutput=1
\usepackage{jheppub}

\usepackage{latexsym,amsbsy}
\usepackage{slashed}

\usepackage[compat=1.1.0]{tikz-feynman}

\newcommand{\BE}{\begin{equation}}
\newcommand{\EE}{\end{equation}}
\newcommand{\BQ}{\begin{equation} \begin{array}{c}}
\newcommand{\EQ}{\end{array}\end{equation}}
\newcommand{\BT}{\begin{theorem}}
\newcommand{\ET}{\end{theorem}}

\newcommand{\bc}{\begin{center}}
\newcommand{\ec}{\end{center}}

\newcommand{\DHS}{\breve{D}}
\newcommand{\FHS}{\breve{F}}
\newcommand{\GHS}{\breve{G}}

\newcommand{\dT}{\widetilde{d}}
\newcommand{\AT}{\widetilde{A}}
\newcommand{\DT}{\widetilde{D}}
\newcommand{\CT}{\widetilde{C}}
\newcommand{\FT}{\widetilde{F}}
\newcommand{\LX}{\Lambda}

\newcommand{\lX}{\lambda}

\title{Chirality, a new key for the definition of the connection and curvature of a Lie-Kac superalgebra}

\author{Jean Thierry-Mieg}

\affiliation{NCBI, National Library of Medicine, National Institute of Health, \\
  8600 Rockville Pike, Bethesda MD20894, U.S.A.}

\emailAdd{mieg@ncbi.nlm.nih.gov}

\abstract{A natural generalization of a Lie algebra connection, or Yang-Mills field,
to the case of a Lie-Kac superalgebra, for example SU(m/n), just in terms of ordinary
complex functions and differentials, is proposed. Using the chirality $\chi$ which defines
the supertrace of the superalgebra:
$STr(...) = Tr (\chi ...)$, we construct a covariant differential:
$D = \chi (d + A) + \Phi$, where A is the standard even Lie-subalgebra connection 1-form
and $\Phi$ a scalar field valued in the odd module. Despite the fact that $\Phi$
is a scalar, $\Phi$ anticommutes with $(\chi A)$ because $\chi$ anticommutes with
the odd generators hidden in $\Phi$. Hence
the curvature $F = DD$ is a superalgebra-valued linear map which respects
the Bianchi identity and correctly defines a chiral parallel transport
compatible with a generic Lie superalgebra structure.}

\begin{document}
\maketitle
\flushbottom

\section {Introduction}

In differential geometry and in Yang-Mills theory,
the  Lie algebra-valued connection $A$
and the curvature 2-form $F = dA + AA$
are fundamental concepts. A natural generalization
to the case of a (simple) Lie-Kac superalgebra \cite{Kac1},
for example $SU(m/n)$, raises a sign issue.
The difficulty stems from the fact that 1-forms naturally anticommute,
so the product $AA$ appearing in the definition of the curvature
$F$ hides a matrix commutator $AA = \frac{1}{2} \;A^a A^b\; [\lX_a,\lX_b]$
which, as desired, closes in a Lie algebra. But in a superalgebra,
we need to associate to the odd generators a commuting field $\Phi$
such that the product $\Phi\Phi$ is symmetric and
hides an anticommutator
$\Phi\Phi = \frac{1}{2}\; \Phi^i\Phi^j\;\{\lX_i,\lX_j\}$.
Yet this commuting scalar 
field $\Phi$ must anticommute with $A$
in order to generate, in the even-odd sector, a matrix commutator
$[\lX_a, \lX_i]$, and $\Phi$ must be odd as an element of the
differential calculus  so that the curvature 2-form $F = DD$ remains
a linear map.

Several constructions have been proposed. They
have in common that they postulate that the scalar fields $\Phi$,
called $L$ in Quillen \cite{Quillen,MQ86}, are odd
relative to the Yang-Mills 1-forms $A$.
The $L$ are often represented as 1-forms relative to an enlarged space
which can be a super-space involving anticommuting Grassman coordinates \cite{Stavra,Dumi,CGP}, or
a discrete space, for example $Z_2$, equipped with discrete differentials \cite{dbv88,dbv90,CL,C1,C2,CQ1,CQ2,CQ3}.
But these odd $L$ fields cannot easily be represented in quantum field theory,
where bosons necessarily commute. Therefore $L$ cannot correspond to the Higgs fields
of the standard model of the fundamental interactions.

In this note, using a slight modification of the
usual formalism of covariant differentials and connections,
we construct a surprisingly simple universal solution to this problem.
Any superalgebra \cite{Kac1} carries a $Z_2$ chirality operator $\chi$ which commutes with
the even generators spanning the Lie subalgebra, anticommutes with the odd generators spanning the odd module,
and defines the supertrace
$STr (...) = Tr (\chi ...)$. Consider a standard 1-form $A$, valued in
the even Lie subalgebra, together with a scalar $\Phi$ valued in the odd module, 
and let us define the new chiral covariant differential $\DT = \chi\; (d+A) + \Phi$.
This algebraic structure implies three direct consequences.
The curvature $\FT = \DT^2$ is valued in the adjoint representation of the
superalgebra. It defines a linear map. It satisfies the Bianchi identity $\DT\FT = 0$.
The proof relies on 3 rules: $\chi A$ anticommutes with $\chi A$ since $A$ is a 1-form;
$\Phi$ commutes with $\Phi$ since $\Phi$ is a scalar;
finally $\chi A$ anticommutes with $\Phi$, despite the fact that $\Phi^i$ is a 
standard commuting complex-valued function, because
$\chi$ anticommutes with the odd generators $\lX_i$ hidden in $\Phi = \Phi^i\lX_i$.

As a result, $\DT$ defines a parallel transport on the super-principal bundle, 
where the base space is an ordinary differential
manifold and the fiber
is isomorphic to a Lie-Kac superalgebra. $\DT$ is defined in terms of standard complex-valued 
functions, without using non standard Grassman coordinates
or discrete differentials. This natural construction may open the way
to a reanalysis of the Ne'eman-Fairlie $SU(2/1)$ superalgebraic model
of the electroweak chiral interactions of 
leptons \cite{N1,F1} and quarks \cite {DJ,NTM1} 

\section{Definition of a superalgebra}

Let us consider a finite dimensional
basic classical Lie-Kac superalgebra \cite{Kac1}.
The superalgebra acts on a $Z_2$ graded finite dimensional 
vector space $V = V_0 + V_1$ over the complex numbers. 
The chirality matrix $\chi$ is diagonal, with eigenvalue $1$ on the
$V_0$ and $-1$ on $V_1$. $\chi$ defines the supertrace
\BE
STr(...) = Tr (\chi\;...) \;.
\EE

Each generator is represented by 
a finite dimensional matrix of complex numbers. Note that 
in our approach, we do not need anticommuting Grassman numbers.
The even generators are denoted $\lX_a$ and the odd generators $\lX_i$.
$\chi$ commutes with the $\lX_a$ and anticommutes with the $\lX_i$
\BE
 [\chi,\;\lX_a] = \{\chi,\;\lX_i\} = 0 \;.
\EE
The $\lX$ matrices close under (anti)-commutation
\BE
  [\lX_a,\;\lX_b] = f^c_{ab} \;\lX_c
\;,\;\;\;
  [\lX_a,\;\lX_i] = f^j_{ai} \;\lX_j
\;,\;\;\;
  \{\lX_i,\;\lX_j\} = d^a_{ij} \;\lX_a
\;,
\EE
and satisfy the super-Jacobi relation with 3 cyclic permuted terms:
\BE
   (-1)^{AC} \{\lX_A, \{\lX_B,\;\lX_C]] + 
   (-1)^{BA} \{\lX_B, \{\lX_C,\;\lX_A]] + 
        (-1)^{CB} \{\lX_C, \{\lX_A,\;\lX_B]]  = 0\;.
\EE
where the mixed bracket denotes either a commutator or an anticommutator as needed.
The quadratic Casimir tensor $(g_{ab},g_{ij})$, also called the super-Killing metric, is defined as
\BQ
  g_{ab} = \frac{1}{2} STr (\lX_a\lX_b)\;,
\\
g_{ij} = \frac{1}{2} STr (\lX_i\lX_j)
\EQ
The even part $g_{ab}$ of the metric is as usual symmetric,
but because the odd generators anticommute with
the chirality hidden in the supertrace (2.1),
its odd part $g_{ij}$ is antisymmetric:
$Tr(\chi\lX_j\lX_i)=-Tr(\lX_j\chi\lX_i)=-Tr(\chi\lX_i\lX_j)$.
The structure constants can be recovered from the
supertrace of a product of 3 matrices
\BQ
f_{abc} = g_{ae}\,f^e_{bc} = \frac{1}{2} STr (\lX_a\,[\lX_b,\lX_c])\;,
\\
f_{iaj} = g_{ik}\,f^k_{aj} = \frac{1}{2} STr (\lX_i\,[\lX_a,\lX_j])\;,
\\
d_{aij} = g_{ae}\,d^e_{ij} = \frac{1}{2} STr (\lX_a\,\{\lX_i,\lX_j\})\;.
\EQ

In an abstract way, a Lie-Kac superalgebra is specified by providing
a set of even and odd generators $\LX^a$ and $\LX^i$ satisfying
closure and the graded Jacobi identity, plus a chirality
operator $X$ commuting/anticommuting with the even/odd generators.
A finite dimensional linear representation $\rho$ over the complex numbers
is defined by providing a set of finite dimensional complex
matrices denoted $\lX^a$ and $\lX^i$, such that $\rho(\LX) = \lX$,
and a grading matrix $\chi = \rho(X)$ satisfying the same relations.
The matrix $\chi$ is not an addition, but an intrinsic part
of the definition of the linear representation $\rho$.
It will play a central role in our new definition of
a superalgebra superconnection.

\section {Lie algebra connections respect the Bianchi identity}

In the Lie algebra case, a connection $A$ is a Lie algebra-valued 1-form
\BE
A = A^a_{\mu}(x) dx^{\mu} \lX_a \;,
\EE
where $A^a_{\mu}(x)$, the Yang-Mills vector-field of the physicists,
is an ordinary commuting complex valued function, and $dx^{\mu}$,
the exterior differentials of the coordinates,
are anticommuting 1-forms.  $A$ can be combined
with the exterior differential $d = \partial_{\mu}dx^{\mu}$ 
to construct a covariant differential
\BE
D = d + A\;.
\EE
Let us iterate the action of $D$ on a vector $\psi$.
Since $d$ satisfies the graded Leibniz rule
\BE
  d (A ...) = (dA) ... - A d(...)
\EE
we find that $D^2$ is tensorial, i.e. defines a linear map
with respect to scalar functions, as $D^2\psi$ is independent of $d\psi$
\BQ
 D^2\,\psi = (d+A)(d+A)\,\psi = (dA + AA)\,\psi + 0\,d\psi
\EQ
allowing us to define the curvature 2-form
\BQ
F = D^2 = dA + AA
\EQ
The second crucial observation is that, because the $A$ are 1-forms,
the product $AA$ is antisymmetric in $(a,b)$
and hence proportional to the commutator of the $\lX$ matrices
\BE
AA = \frac{1}{2}  \; A^a_{\mu}A^b_{\nu}\;dx^{\mu}dx^{\nu}\; [\lX_a,\lX_b]
\EE
and since the commutator closes in the algebra (2.3), the curvature 2-form
F is valued in the adjoint representation of the Lie algebra
\BE
F = F^a\,\lX_a = (dA^a + \frac{1}{2}f^a_{bc} A^bA^c)\;\lX_a\;.
\EE
Let us now verify that the triple action of $D$ is associative:
\BE
  D(DD)\psi = (DD)D\psi \;.
\EE
This condition is equivalent to the Bianchi identity
\BE
 DF = dF + AF - FA = 0 \;.
\EE
In this equation, the terms linear in $dA$, coming from $F = dA + AA$,
cancel out and we are left with the constraint
\BE
A^b A^c A^d\;f^a_{be}f^e_{cd} = 0\;.
\EE 
Because the $A$ 1-forms anticommute, this constraint (3.10) is
satisfied thanks to the Jacobi identity (2.4). Thus, for a Lie algebra-valued
connection 1-form, the Bianchi identity (3.9) is satisfied and the covariant differential
$D$ is associative (3.8).

\section{About superconnections}

Let us now try to extend those definitions to a Lie-Kac superalgebra.
In his seminal paper on superconnections, Quillen \cite{Quillen,MQ86} explains that
if we define the covariant differential as
\BE
\DT = \dT + \AT
\EE
then the connection form $\AT$ must be odd with respect to the
differential calculus, meaning that
$\dT^2$ must vanish and $\dT$ must satisfy the graded Leibniz rule (3.3).
As shown in the previous section, these rules are naturally valid for the
exterior differential $d$ and the 1-forms $A$, 
if $A$ is a Lie algebra valued connection. The question is to find
a generalization if $\AT$ is valued in the adjoint representation
of a Lie-Kac superalgebra. 

A first naive guess would be to construct the superconnection as a 1-form
\BE
D = d + A = dx^{\mu}\;(\partial_{\mu} + A^a_{\mu}\lX_a + A^i_{\mu}\lX_i)\;.
\EE 
However, this definition would be inconsistent because the curvature $F$
would involve the commutator of the odd matrices
\BE
F = dA + AA =  ... +  A^i_{\mu}A^j_{\nu}\;dx^{\mu}dx^{\nu}\; \frac{1}{2} [\lX_i,\lX_j]
\EE
which does not close on the even matrices.

The next possibility is to define the theory over superspace \cite {Stavra,Dumi,CGP}. 
This method introduces a lot of subtleties.
As usual in supersymmetric theories, each component field $A(x,\theta)$ 
can be developed as a finite polynomial
over the Grassman coordinates $\theta$, implicitly
generating a great number of auxiliary fields. 
But the exterior differentials  $d\theta$ of the Grassman
coordinates generate an even larger complexity. 
The $d\theta$ commute, so polynomials of arbitrary degree in $d\theta$ can exist.
But in a way, this approach cannot solve our original question.
Indeed, the intrinsic geometry of Elie Cartan, 
which deals with exterior differential and exterior forms,
is by construction independent of the choice of coordinates on the
base space, which can be a standard differential manifold $(x)$,
or a superspace $(x,\theta)$. For example, the  Lie algebra covariant differential
over superspace simply reads 
$D = d + A(x,\theta)$ where $d=\partial_{\mu}dx^{\mu} + \partial_{\theta}d\theta$.
It is therefore apparent that the introduction of a superspace
does not resolve the paradoxical sign rules that we described in the introduction,
unless we restrict the even components of the connection $A^a$ to depend only on $dx$
and the odd components $A^i$ only on $d\theta$: $A = A^a_{\mu} \lX_a dx^{\mu} + A^i \lX_i d\theta$.
But such an assumption contradicts the essence of Cartan's intrinsic calculus
since this constraint is not invariant under a generic super-rotation of the coordinates.

In 1982, with Ne'eman \cite{NTM82}, we introduced a superconnection as the odd part of
the de Rham complex of forms of all degrees, valued in a Lie superalgebra.
\BE
\AT = \Phi + A + B + ... = \Phi^i\lX_i + A^a\lX_a + B^i\lX_i +...
\EE
where $\Phi$ is a scalar field valued in the odd module of the superalgebra,
$A$ is a 1-form valued in the even subalgebra, $B$ a 2-form valued in the 
odd module and so on. In this method, when we compute $(d+\Phi+A)^2$, the term in $AA$
gives as usual the commutator of the even generators (3.6) and 
since $\Phi$ is a scalar, the term in $\Phi\Phi$ is symmetric and generates
as desired the anticommutator of the odd generators
\BE
F = d(A+\Phi) + (A+\Phi)(A+\Phi) = ... +  \Phi^i\Phi^j\; \frac{1}{2} \{\lX_i,\lX_j\}
\EE
However, $d(\Phi ...)$ does not obey the graded Leibniz rule (3.3) but gives a plus sign
which breaks the tensorial nature of the curvature $F$
\BE
  DD \psi = ... + d(\Phi\,\psi) + \Phi \,d\psi = ... + (d\Phi)\, \psi + 2 \Phi\, d\psi
\EE
which now depends on $d\psi$. At the same time, this proposition does not generate
in the even-odd sector the desired commutator $[\lX_a,\lX_i]$, but an anticommutator
\BE
A\Phi + \Phi A = A^a\Phi^i (\lX_a\lX_i + \lX_i\lX_a)
\EE 
which does not close in the superalgebra. The same problem affects
the product $AB$ and all other terms in the even-odd sector.

In 1985, Quillen \cite{Quillen,MQ86} proposed a general construction
applicable to  non-trivial bundles
and resolved this problem by adding to the Yang-Mills covariant differential
$D = d + A$ a new $L$ term
\BE
D = d + A + L
\EE
where $L$ is valued in the odd module of the superalgebra $L = L^i\lX_i$,
and $L^i$ is by definition odd with respect to the differential calculus,
meaning that $L$ anticommutes with $A$ and satisfies the graded Leibniz rule.
In essence, Quillen postulates that $L$ takes
values in another algebra which anticommutes with the exterior differentials.
This method was later adopted by Ne'eman and Sternberg who
postulate that $\omega_0 L_{01} =-L_{01} \omega_0$, 
see equation 1.3ab of \cite{NS90} or section 5.8 of \cite{NSF}. 
This is a correct mathematical construction, but it is not
applicable to physics. The limitation
resides in its transposition to quantum field theory (QFT).
The connection 1-forms $A = A^a_{\mu}\lX_a dx^{\mu}$ are naturally represented by
the Yang-Mills vector bosons $A^a_{\mu}$, the even connection $L = L^i\lX_i$ 
looks like a scalar fields $L^i$ which, in the $SU(2/1)$ case, has the quantum number
of a Higgs Boson. Unfortunately bosons commute. Therefore there is no clear way
to represent in QFT the required anticommutativity of $A^a_{\mu}$ and $L^i$.
This is probably why, up to now, the literature on the superalgebraic $SU(2/1)$ model
of the weak interactions \cite{CL,C1,CQ1,CQ2,CQ3,N1,F1,DJ,NTM1,NTM82,NS90}
does not go beyond a classical analysis and
never mentions renormalization and the Feynman diagrams of QFT.

\section {New definition of a chiral superconnection satisfying the Bianchi identity}

Let us now introduce a very simple new
definition which naturally applies to any Lie-Kac superalgebra.
Consider the exterior differential
\BQ
\DT = \dT + \AT \;,
\\
\dT = \chi\,d\;,\;\;\AT = \chi A + \Phi\;,
\\
\DT = \chi\, dx^{\mu}\;(\partial_{\mu} + A^a_{\mu}\lX_a) + \Phi^i\lX_i\;,
\EQ
where $\chi$ is the chirality operator instrumental in the definition of the supertrace (2.1) and
$D$ is the standard Lie algebra covariant differential $D = d + A$ (3.2).
This definition is valid for any superalgebra, since they are all equipped
with a chirality operator. It does not involve Grassman numbers.
$A^a_{\mu}(x)$ and $\Phi^i(x)$ are standard complex-valued fields, respecting
the spin-statistics theorem. 
As in equation (3.6), the $\FT =...+ AA$ term generates
the desired commutator of the even $\lX_a$ generators.
As in equation (4.5), the $\FT =...+ \Phi\Phi$ term generates
the desired anticommutator of the odd $\lX_i$ generators.
The new result is that $\chi d$ and $\chi A$, which are 1-forms,
anticommute with the scalar field $\Phi$, not because of their exterior degree, 
but because the odd matrices $\lX_i$, hidden in $\Phi = \Phi^i\lX_i$, anticommute 
(2.2) with the chirality operator
$\chi$  which defines the supertrace (2.1).
As a consequence, $(\dT = \chi d)$ satisfies the graded Leibniz rule (3.3)
\BE
  \dT (\AT ...) = (\dT\AT) ... - \AT \dT(...)\;,
\EE
the sign problem of equation (4.6) disappears and the curvature 
2-form $\FT$ defines a linear map as $\DT\DT\Psi$ does not depend on $d\Psi$.
Moreover, since $\chi A$ anticommutes with $\Phi$, 
all the commutators and anticommutators have the
desired signs and $\FT$ is valued in the adjoint representation of the superalgebra :
\BQ
\DT\DT\,\psi = \FT \psi = (\FHS + \GHS) \psi \;,
\\
\FHS = F + \Phi\Phi\;,\;\;\;\GHS = \DHS\Phi\;,
\EQ
where the covariant differential $\DHS\Phi$ contains the usual Lie algebra even-odd commutator
\BE
\DHS\Phi = \chi d\Phi + \chi A \Phi + \Phi \chi A = \chi (d\Phi + A\Phi - \Phi A)  \;.
\EE
Please observe how rule (2.2) defining the gradation of a superalgebra
implies that $\Phi \chi = - \chi \Phi$
thus changing the wrong anticommutator of equation (4.7) into the desired commutator (5.4).
The new covariant differential $\DT$ is associative 
\BE
\DT(\DT\DT) \psi = (\DT\DT)\DT \psi
\EE
because the Bianchi identity
\BE
\DT\FT = \dT\FT + \AT\FT - \FT\AT = 0
\EE
is satisfied
thanks to the super-Jacobi identity (2.4) since all the terms trilinear in $(A,\Phi)$ appear with the
proper combinations of signs.

It may seem that an alternative construction: $D = \chi\DT = d + A + \chi \Phi$, closer
to \cite{Quillen,NSF}, could also work, but this is not the case. Since $d$ commutes with $\chi$,
$\chi \Phi$ is even relative to the differentials $d$. Thus, the term
in $\chi \Phi \, d\psi$ does not cancel out and the curvature does not define a linear map.
This problem subsists for any definition of the type $D = d + \Phi^i\mu_i$ where $\Phi^i$ is
a commuting scalar and $\mu_i$ an arbitrary matrix of complex numbers. The cornerstone
of our new consistent definition is to join the chirality to the exterior differential $\dT = \chi d$
so that $\Phi$ becomes odd relative to $\dT$. The definition of Quillen 
$D = d + A + L$ is consistent because he assumes that $L$ is valued in
an additional graded algebra odd relative to $d$,
but then $L$ cannot be represented in quantum field theory by a commuting scalar Boson
and cannot be interpreted as the Higgs field of the standard model. We circumvent
this problem by defining $\Phi$ as a standard commuting complex function and
decorating the differential by the chirality matrix $\dT = \chi d$. Following
Occam, this definition is also more economic than Quillen's because $\chi$ is not
an additional operator but a constitutive element of the definition of a linear
representation of a Lie-Kac superalgebra.

Please notice that in these equations $\psi$ does not denote a space-time spinor but an element
of a generic linear representation of the superalgebra. Similarly, $\chi$ acts on $\psi$
and anticommutes with the odd generators of the superalgebra but is not necessarily related to
the $\gamma_5$ chirality operator of the spinor space.
The actual pairing of $\chi$ and $\gamma_5$
as a representation of the $CP$ invariance of the weak interactions is discussed in
\cite{TM20}.

\section {Application to Quantum Field Theory}

The distinction between the even and odd
generators of a superalgebra is intrinsic, they commute or anticommute with
the chirality operator defining the supertrace (2.2). However, the carrier space
$V = V_0 + V_1$ is intrinsically symmetric. Presenting it as a $Z_2$ graded space
is in a way misleading because the two parts stand on equal footing. It is split,
but it does not carry the distinct nature of $0$ and $1$ under addition and multiplication.
There is no good reason to associate $V_0$ to bosons and $V_1$ to fermions, or vice-versa.
For example, the quadratic super-Casimir operator is a multiple of the identity
with the same eigenvalue $k$ on $V_0$ and on $V_1$. It is more natural to
think of $\chi$ as the chirality and to think of a superalgebra as mapping
left to right fermions and vice versa \cite{N1,F1,CL,NSF}, rather than bosons to fermions \cite{CNS}.

Consider now the Dirac equation associated to our new chiral superconnection, 
\BE
(\chi (\partial_{\mu} + A_{\mu}) \gamma^{\mu} + \Phi)\psi = 0 \;.
\EE
The presence of the chirality operator in front of the space-time derivative $\partial_{\mu}$
solves the important problem of the sign of the energy of the vector Boson $A^a_{\mu}$.
In a superalgebra, the natural normalization of the even matrices involves the
supertrace
\BE
STr (\lX_a\lX_b) = Tr (\chi \lX_a\lX_b) = \pm 2 \;\delta_{ab} \;.
\EE
so that in a gauge theory of the simple superalgebra $SU(m/n)$, either the $SU(m)$ or the $SU(n)$ matrices
have a negative norm. Hence there is a risk of constructing states with a negative energy. 
But using $\DT = \chi (d + A)$, the sign of the time derivative 
in the Dirac equation is switched when we flip the sign of the chirality,
maintaining the sign of the product energy x time. This restores
the symmetry between $SU(m)$ and $SU(n)$. 
We conclude that both sectors, $SU(m)$ and $SU(n)$ follow the usual axioms and
respect unitarity.

Note that in his many presentations, for example  \cite{CL,C2},
Alain Connes emphasizes the importance of the chirality in the
definition of his Dirac-Yukawa operator. But although his analysis is closely
related to the chiral differential $\dT = \chi d$ defined in equation (5.1),
Connes in his works concerning the standard model of the fundamental interactions does not
use simple superalgebras and does not make the connection to the $SU(2/1)$
of Ne'eman \cite{N1} and Fairlie \cite{F1}. Maybe, our new formalism will help.

\section {Chern-Simons Lagrangian}

The standard construction of the Chern-Simons Lagrangian can be generalized to the
chiral super connection. Consider the exterior form
\BE
 \CT = \AT \dT \AT + \frac {2}{3} \;\AT\,\AT\,\AT\;.
\EE
By taking the exterior differential we recover the usual formula
\BE
 STr (\dT \CT) = STr (\FT\,\FT) = g_{ab} \FHS^a \FHS^b + g_{ij} D\Phi^i\,D\Phi^j\;, 
\EE
where $(g_{ab}, g_{ij})$ is the super-Killing metric of the superalgebra (2.5) and $\FHS$ is defined in (5.3).
Observe the natural occurrence of the supertrace, needed to recover the super-Killing metric.
As usual, the term in $A^4$ present in $\FT^2$ but not in $d\CT$ is eliminated
by the Jacobi identity when we sum over the cyclic permutations of the indices $(def)$
\BE
\Sigma_{def} \;g_{ab} \,f^a_{cd}\,f^b_{ef} = \Sigma_{def} \; f_{cda} \,f^a_{ef} = 0\;.
\EE
In $\FHS^2$, we also have a term in $\Phi^4$ which is eliminated by the super-Jacobi identity
\BE
\Sigma_{jkl} \;g_{ab} \,d^a_{ij}\,d^b_{kl} = \Sigma_{jkl} \; f_{ija} \,d^a_{kl} = 0\;.
\EE
The topological Lagrangian $STr (\FT\,\FT)$ is as usual closed and locally exact. The
unusual feature is that it is not a homogeneous 4-form but
as in \cite{Quillen,MQ86} a differential exterior form of mixed degree, which interferes
with the normal way in which we compute the Chern classes.

\section{Discussion}

Despite the importance of intrinsic differential geometry, and despite the abundance
of textbooks and research devoted to superconnections (see for example \cite{MS,Dumi,CGP}),
we show in this paper that the construction
of a connection valued in a Lie-Kac superalgebra \cite{Kac1} defined in terms of
ordinary matrices, complex valued functions and ordinary differential forms can still be simplified. 
The method presented here differs from previous ones \cite{Quillen,CL} by a slight modification:
in order to generate the correct signs in the definition of the curvature
and hence recover the Bianchi identity as a consequence of the super-Jacobi identity,
we have transferred in our definition $\DT = \chi \,(d + A) + \Phi$ 
the commutation rule burden from the odd to the even generators. 
The odd field $\Phi^i$, which carries the odd indices, is a normal commuting scalar field.
However, the even field $\chi\,A^a_{\mu}$ carries a chirality operator $\chi$
which is an intrinsic element of any linear representation of
a Lie-Kac superalgebra and which anticommutes with the odd generators (2.1). With this simple change,
all the equations are explicitly covariant and are written in term
of ordinary complex valued functions. This may open the way to new 
applications in geometry and quantum-field theory \cite{TM20}.

\section*{Acknowledgments}
This research was supported by the Intramural Research Program of the National Library of Medicine, National Institute of Health.
We are extremely grateful to the referee for his patience and his illuminating explanations of the literature
which had a profound impact on the manuscript. We also thank 
Danielle Thierry-Mieg for clarifying the ideas and the presentation.


\end{document}